\documentclass{elsart3}
\usepackage{graphicx}
\usepackage{amsmath}
\usepackage{amssymb}

\begin{document}

\begin{frontmatter}

\title{
Decoherence of Phase Qubit using High-$T_c$ Superconductor
}

\author[address1,address2]{Shiro~Kawabata\thanksref{thank1}},
\author[address3]{Satoshi~Kashiwaya}
\author[address4]{Yasuhiro~Asano} 
and
\author[address5]{Yukio~Tanaka}

\address[address1]{Nanotechnology Research Institute and SYNAF, National Institute of Advanced Industrial Science and Technology (AIST), 1-1-1 Umezono, Tsukuba, Ibaraki 305-8568, Japan}

\address[address2]{International Project Center for Integrated Research on Quantum Information and Life Science (MILq Project), Hiroshima University, Higashi-Hiroshima, 739-8521, Japan}

\address[address3]{Nanoelectronics Research Institute, AIST, 1-1-1 Umezono, Tsukuba, Ibaraki 305-8568, Japan}

\address[address4]{Department of Applied Physics, Hokkaido University, Sapporo 060-8628, Japan}

\address[address5]{Department of Applied Physics, Nagoya University, Nagoya 464-8603, Japan}

\thanks[thank1]{
Corresponding author. 
E-mail: s-kawabata@aist.go.jp}

\begin{abstract}
We discuss how to make use of high-$T_c$ $d$-wave Josephson junctions in the construction of a phase qubit.
We especially focus on the effect of the quasiparticle dissipation and the zero energy bound state on the macroscopic quantum tunneling which corresponds to the final measurement process of the $d$-wave phase qubit.
\end{abstract}

\end{frontmatter}

\section{Introduction}
Since macroscopic systems are inherently dissipative, there arises a fundamental question of what is the effect of dissipation on the macroscopic quantum tunneling (MQT).
This issue was solved by Caldeira and Leggett by using the path-integral method and they showed that the MQT is depressed by dissipation~\cite{rf:CL81}.
This effect has been verified in experiments on $s$-wave Josephson junctions shunted by an Ohmic normal resistance)~\cite{rf:Cleland88}.
As was mentioned by Eckern {\it et al.}, the influence of the quasiparticle dissipation is quantitatively weaker than that of the Ohmic dissipation on the shunt resistor~\cite{rf:Eckern84}.
This is due to the existence of an energy gap $\Delta$ for the quasiparticle excitation in superconductors.
Therefore, in an ideal $s$-wave Josephson junction without the shunt resistance, the suppression of the MQT rate due to the quasiparticle dissipation is very weak.

In this paper, we will discuss how to make use of high-$T_c$ $d$-wave Josephson junctions (Fig. 1) in the construction of a phase qubit.
In the $d$-wave superconductors, the the gap vanishes in certain directions~\cite{rf:d-wave1}, hence quasiparticles can be excited even at sufficiently low temperature regime.
Moreover in the case of $d$-wave junctions along $ab$-plane (Fig. 1(b)), the zero energy bound states (ZES) are generated by the combined effect of the Andreev reflections and the sign change of the $d$-wave order parameter symmetry~\cite{rf:d-wave1}.
The ZES are bound states for the quasiparticle at the Fermi energy.
Therefore the ZES may gives rise to a strong dissipation.
Below we will show a theoretical proposal of the $d$-wave phase qubit and the theoretical analysis of the effect of quasiparticle decoherence (dissipation) on the MQT which corresponds to the final measurement process of the $d$-wave phase qubit.
Note that recently the influence  of the quasiparticle decoherence for another type of $d$-wave qubit ($quiet$ qubit) have been discussed by Amin $et$ $al.$~\cite{rf:Amin} and Fominov $et$ $al$~\cite{rf:Fominov}.

\section{$d$-wave phase qubit}

The $d$-wave phase qubit is constructed from the $d$-wave Josephson junction with a fixed dc current source (Fig. 1).
Like the flux qubit~\cite{rf:Review}, this type of qubit is insensitive to the effect of the background charge fluctuations.

\begin{figure}[t]
\includegraphics[width=6.0cm]{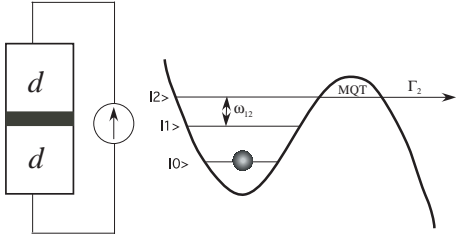}
\caption{Schematic drawing of the $d$-wave phase qubit using the current biased $d$-wave Josephson junctions. }
\label{fig1}
\end{figure}

The ground state $|0 \rangle$ and the first excited state $|1 \rangle$ in a metastable potential are used as the qubit.
Rabi oscillations between the stats $|0 \rangle$ and $|1 \rangle$ can be observed by irradiating the qubit with microwaves at a frequency $\omega_{10}$ and then measuring the occupation probability of being in the state $|1 \rangle$.
The measurement of the qubit state utilizes the escape from the cubic potential via MQT. To measure the occupation probability $P_1$ of state $| 1 \rangle$, we pulse microwaves at frequency $\omega_{12}$, driving a $1\to 2$ transition. The
large tunneling rate $\Gamma_2$ then causes state $| 2 \rangle$ to rapidly
tunnel. 
After tunneling, the junction behaves as an open
circuit, and a dc voltage of the order of the superconducting
gap appears across the junction. 
Thus the occupation probability $P_1$ is equal to
the probability of observing a voltage across the junction
after the measurement pulse.
\section{Effect of nodal quasiparticle}
\begin{figure}[b]
\includegraphics[width=5.0cm]{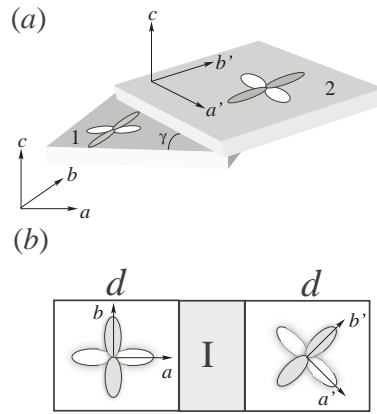}
\caption{Schematics of (a) the $c$-axis Josephson junction and (b) the $d$-wave Josephson junction along the $ab$-plane.}
\label{fig2}
\end{figure}

We will show the calculation of the MQT rate for the $d$-wave phase qubit using the  $c$-axis Josephson junction~\cite{rf:Takano02} (Fig. 2(a)).
In the $c$-axis Josephson junctions, the ZES is completely absent~\cite{rf:d-wave1}.
In Fig. 1(a), $\gamma$ is the twist angle about the $c$-axis ($0 \le \gamma \le \pi/4$). 
Such a twist junction was recently fabricated by using the single crystal whisker of Bi${}_2$Sr${}_2$CaCu${}_2$O${}_{8+\delta}$.
Takano $et$ $al.$ have measured the twist angle dependence of the $c$-axis Josephson critical current and showed a clear evidence of the $d_{x^2-y^2}$ symmetry of the pair potential~\cite{rf:Takano02}.
In the following, we assume that the tunneling between the two superconductors is described in terms of the coherent tunneling ($\left| t(\mbox{\boldmath $k$},\mbox{\boldmath $k$}')\right|^2 
 = 
 \left| t \right|^2 \delta_{\mbox{\boldmath $k$}_{\parallel},\mbox{\boldmath $k$'}_{\parallel}}
$).
For simplicity, we also assume that each superconductors consist of single  CuO${}_2$ layer, $\Delta_1 (\mbox{\boldmath $k$})=\Delta_0 \cos 2 \theta$, and $\Delta_2 (\mbox{\boldmath $k$})=\Delta_0 \cos 2 \left( \theta + \gamma \right)$.
Moreover, we consider the low temperature limit ($k_B T \ll \Delta_0$).

By using the functional integral method, the ground partition function for the system can be written as follows
\begin{eqnarray}
Z
= 
\int 
D \phi (\tau) 
\exp
\left[
  - \frac{S_{\mathrm{eff}}[\phi]}{\hbar} 
\right]
,
\end{eqnarray}
where $\phi=\phi_1-\phi_2$ is the phase difference across the junction.
In this equation, the effective action is given by 
\begin{eqnarray}
S_{\mathrm{eff}}[\phi]
&= &
\int_{0}^{\hbar \beta} d \tau 
\left[
   \frac{M}{2} 
   \left(
   \frac{\partial \phi ( \tau) }{\partial \tau}
   \right)^2
   + 
   U(\phi)
\right]
+
S^{[\alpha]}[\phi]
,
\nonumber\\
\\
S^{[\alpha]}[\phi]
&= &
-
\int_{0}^{\hbar \beta}  d \tau 
 \int_{0}^{\hbar \beta} d \tau'
  \alpha (\tau - \tau') \cos \frac{\phi(\tau) - \phi (\tau') }{2}
  ,\nonumber\\
\end{eqnarray}
where $M=C(\hbar/2e)^2$ is the mass ($C$ is the capacitance of the junction) and $ U(\phi) $ is the tiled washboard potential
\begin{eqnarray}
 U(\phi) 
 = 
  -E_J(\gamma) \left(  \cos \phi +  \frac{I_{\mathrm{ext}}}{I_C(\gamma) } \phi \right)
  .
\end{eqnarray}
In this equation, $E_J=\left(  \hbar/2 e \right) I_C$ is the Josephson coupling energy, 
$I_C$ is the Josephson critical current, and $I_{\mathrm{ext}}$ is the external current.

In the following, we will consider the effect of the nodal quasiparticles on the MQT.
For this purpose, we first calculate the dissipation kernel $\alpha (\tau)$ for two types of the $c$-axis junction, $i.e$., (1) $\gamma=0$ and (2) $\gamma \ne 0$ (here we will show the result for $\gamma = \pi/8$ case only.).

In the case of the $c$-axis junction with $\gamma=0$, the nodes of the pair potential in the two superconductors are in the same direction.
Therefore, the node-to-node quasiparticle tunneling is possible even at very low temperatures.
In this case, the asymptotic form of the dissipation kernel at the zero temperature is given by 
\begin{eqnarray}
\alpha(\tau) 
\approx
\frac{3 \hbar^2  | t|^2 N_0^2 }{16 \Delta_0}  \frac{1}{|\tau|^3}
\end{eqnarray}
for $\Delta_0 |\tau| /\hbar \gg1$.
This gives the super-Ohmic dissipation which is qualitatively very weak in compared with the Ohmic dissipation.

On the other hand, in the case of the finite twist angle ($\gamma=\pi/8$), the asymptotic behavior of the dissipation kernel is given by an exponential function due to the suppression of the node-to-node quasiparticle tunneling, $i.e.,$
\begin{eqnarray}
\alpha(\tau)
\sim
 \exp \left( - \frac{1}{\sqrt{2}} \frac{\Delta_0 |\tau| }{\hbar} \right)
\end{eqnarray}
for $\Delta_0 |\tau| /\hbar \gg1$.

The MQT rate at the zero temperature is given by
$
\Gamma
=
\lim_{\beta \to \infty} (2/\beta) \mbox{ Im}\ln Z
\approx
A
\exp \left( -S_B / \hbar \right)
,$
where $S_B= S_{\mathrm{eff}}[\phi_B]$ is the bounce exponent, that is the value of the the action $S_{\mathrm{eff}}$ evaluated along the bounce trajectory $\phi_B(\tau)$.
Using the instanton method, we obtain the analytic expressions for the MQT rate~\cite{rf:Kawabata1,rf:Kawabata2}:
\begin{eqnarray}
\thinspace
\frac{\Gamma(0)}{\Gamma_0(0)}
&\approx&
 \exp \left[ 
 -B(0)
 - 0.14
  \frac{ \hbar I_C(0)}{\Delta_0^2}
  \sqrt{ \frac{\hbar}{2 e} \frac{I_C(0)}{C}}
  \right.
  \nonumber\\
 && 
  \left.
\times
 \left\{
  1 -
  \left(
  \frac{I_{\mathrm{ext}}}{I_C(0)}
  \right)^{2}
\right\}^{5/4}
  \right]
,
\\
\frac{\Gamma(\pi/8)}{\Gamma_0(\pi/8)}
&\approx&
 \exp 
 \left[
    - B\left(\pi/8 \right)
\right]
,
\nonumber\\
\end{eqnarray}
where,
\begin{eqnarray}
B(\gamma)
&=&
\frac{12}{5e} \sqrt{ \frac{\hbar}{2 e} I_C(\gamma) C}
  \left(
                \sqrt{1+ \frac{\delta M(\gamma)}{M} } -1
    \right)
    \nonumber\\
    &&
    \times
	\left\{
1-
    \left(
              \frac{I_{\mathrm{ext}}}{I_C(\gamma)}
    \right)^{2}
	\right\}^{5/4}    ,
\end{eqnarray}
and $\Gamma_0(\gamma)$ is the decay rate without the quasiparticle dissipation.
In eq.(9), $\delta M(\gamma)$ is the renormalization mass.
As an example, for $\Delta_0=42.0$ meV, $I_c(\gamma=0)=1.45 \times10^{-4}$ A, $C=10$ fF, and $I_{\mathrm{ext}}/I_C(\gamma) =0.9$, we obtain
\begin{eqnarray}
\frac{\Gamma(\gamma)}{\Gamma_0(\gamma)}
\approx
\left\{
\begin{array}{rl}
90 \ \% & \quad \mbox{for} \quad \gamma=0 \\
96 \ \% & \quad\mbox{for} \quad \gamma=\pi/8
\end{array}
\right.
.
\end{eqnarray}
\section{Effect of ZES}
In this section we will discuss the influence of the ZES on the MQT in the $d$-wave junctions along the $ab$-plane  ($e.g.,$ YBCO/PBCO/YBCO ramp-edge junctions~\cite{rf:YBCOJJ1} and YBCO grain boundary junctions~\cite{rf:YBCOJJ2}).
In such junctions, the ZES  give a crucial contribution to the Josephson and the quasi-particle current.

In the case of the $d_0/\mbox{I}/d_0$ junction, the node-to-node quasi particle tunneling can contribute to the dissipative quasiparticle current.
However, the ZES is completely absent~\cite{rf:d-wave1}.
Therefore the system shows the super-Ohmic dissipation as in the case of the zero-twist angle $c$-axis junctions or the intrinsic junctions. 
On the other hand, in the case of the $d_0/\mbox{I}/d_{\pi/4}$ junction (Fig. 2(b)), we find that the dissipation kernel is given by 
\begin{eqnarray}
\alpha(\tau) 
\approx
\frac{3 }{16 \sqrt{2}}\frac{\hbar R_Q}{R_N}
\frac{1}{\tau^2}
,
\end{eqnarray}
where $R_Q=h/4e^2$ is the resistance quantum and  $R_N$ is the normal state resistance of the junction.
Therefore the system shows the strong Ohmic dissipation due to the ZES. 
The MQT rate of the  $d_0/\mbox{I}/d_{\pi/4}$ junction is given by~\cite{rf:Kawabata3} 
\begin{eqnarray}
\frac{\Gamma}{\Gamma_0}
\approx
     \exp 
    \left[
   -      \frac{81  \zeta (3)}{32 \sqrt{2} \pi^2 } \frac{R_Q}{R_N}
  \left\{ 1- \left( \frac{I_{ext}}{I_C} \right)^2 \right\}
    \right]
    .
\end{eqnarray}
 For a mesoscopic bicrystal YBCO Josephson junction~\cite{rf:mesoYBCO} ($\Delta_0=17.8 \ $meV, $C=20 \times 10^{-15}\ $F, $R_N = 100  \ \Omega$A$x =0.95$), the MQT rate is estimated as $\Gamma/\Gamma_0 \approx 25\%$.
Therefore the MQT rate in  the $d_0/\mbox{I}/d_{\pi/4}$ junction is strongly suppressed in compared with the $d_0/\mbox{I}/d_{0}$ and the $c$-axis junctions.

\section{Summary}
To summarize, we have investigated the effect of quasiparticle dissipation in the readout process for the $d$-wave phase qubit.
We find the super-Ohmic dissipation in the case of the zero-twist angle $c$-axis junction and the $d_0/\mbox{I}/d_0$ junction.
This dissipation is caused by the node-to-node quasiparticle tunneling between the two superconductors.
Therefore, in this case, the suppression of MQT is very weak.
On the other hand, in the case of $d_0/\mbox{I}/d_{\pi/4}$ junction, the suppression of the MQT rate is very strong in compared with the $c$-axis and the $d_0/\mbox{I}/d_{0}$ junctions due to the appearance of the ZES.
Therefore, it is desirable to use the $d_0/\mbox{I}/d_{0}$ or the $c$-axis Josephson junctions as the phase qubit in order to avoid the strong Ohmic dissipation.
   However the effect of the ZES can be abated by several mechanisms (e.g., by applying magnetic field or by a disorder in the interface)~\cite{rf:d-wave1}.
 Therefore it is interesting to investigate the MQT of the $d_{0}/d_{\pi/4}$ junction in such situations.

 Finally, we would like to comment about recent experimental researches.
 Recently, Inomata $et$ $al.$~\cite{rf:Inomata} and Bauch  $et$ $al.$~\cite{rf:Bauch} have succeeded to observe the MQT in the Bi2212 intrinsic junction and the YBCO grain boundary bi-epitaxial Josephson junction, respectively.



\end{document}